\documentclass[acmtog]{acmart} 
\acmSubmissionID{309}
\pdfoutput=1 

\usepackage{booktabs} 
\usepackage{amsmath}
\usepackage{subfig}
\usepackage{tabularx}
\usepackage{xcolor}
\usepackage{algpseudocode}
\usepackage{algorithm}
\usepackage{caption}

\definecolor{lightblue}{rgb}{0.63,0.78,0.98}

\DeclareMathOperator*{\argmin}{arg\,min}

\newcommand{\sm}[1]{\scriptsize{#1}}
\newcommand{\bd}[1]{\textbf{#1}}

\newcommand{\imagecredits}[1]{%
  \footnotesize{Image credits: #1}
}

\setcopyright{rightsretained} 
\acmJournal{TOG}
\acmYear{2019}
\acmVolume{38}
\acmNumber{4}
\acmArticle{0}
\acmMonth{7} 
\acmDOI{10.1145/3306346.3322993}

\begin{document}
\title{TileGAN: Synthesis of Large-Scale Non-Homogeneous Textures}

\author{Anna Fr{\"u}hst{\"u}ck}
\affiliation{
	\institution{KAUST}
}
\email{anna.fruehstueck@kaust.edu.sa}

\author{Ibraheem Alhashim}
\affiliation{
	\institution{KAUST}
}
\email{ibraheem.alhashim@kaust.edu.sa}

\author{Peter Wonka}
\affiliation{
	\address{ --- Bldg 1, Al Khawarizmi, 4700 KAUST, Thuwal 23955-6900, Kingdom of Saudi Arabia}
	\institution{KAUST}
}
\email{pwonka@gmail.com}

\begin{abstract}
We tackle the problem of texture synthesis in the setting where many input images are given and a large-scale output is required. We build on recent generative adversarial networks and propose two extensions in this paper. First, we propose an algorithm to combine outputs of GANs trained on a smaller resolution to produce a large-scale plausible texture map with virtually no boundary artifacts. Second, we propose a user interface to enable artistic control. Our quantitative and qualitative results showcase the generation of synthesized high-resolution maps consisting of up to hundreds of megapixels as a case in point.
\end{abstract}

%
%

\ccsdesc[500]{Computing methodologies~Computer graphics}
\ccsdesc[500]{Computing methodologies~Texturing}
%
%

\keywords{Texture Synthesis, Image Generation, Deep Learning, Generative Adversarial Networks}

\begin{teaserfigure}
  \centering
  \includegraphics[width=\linewidth]{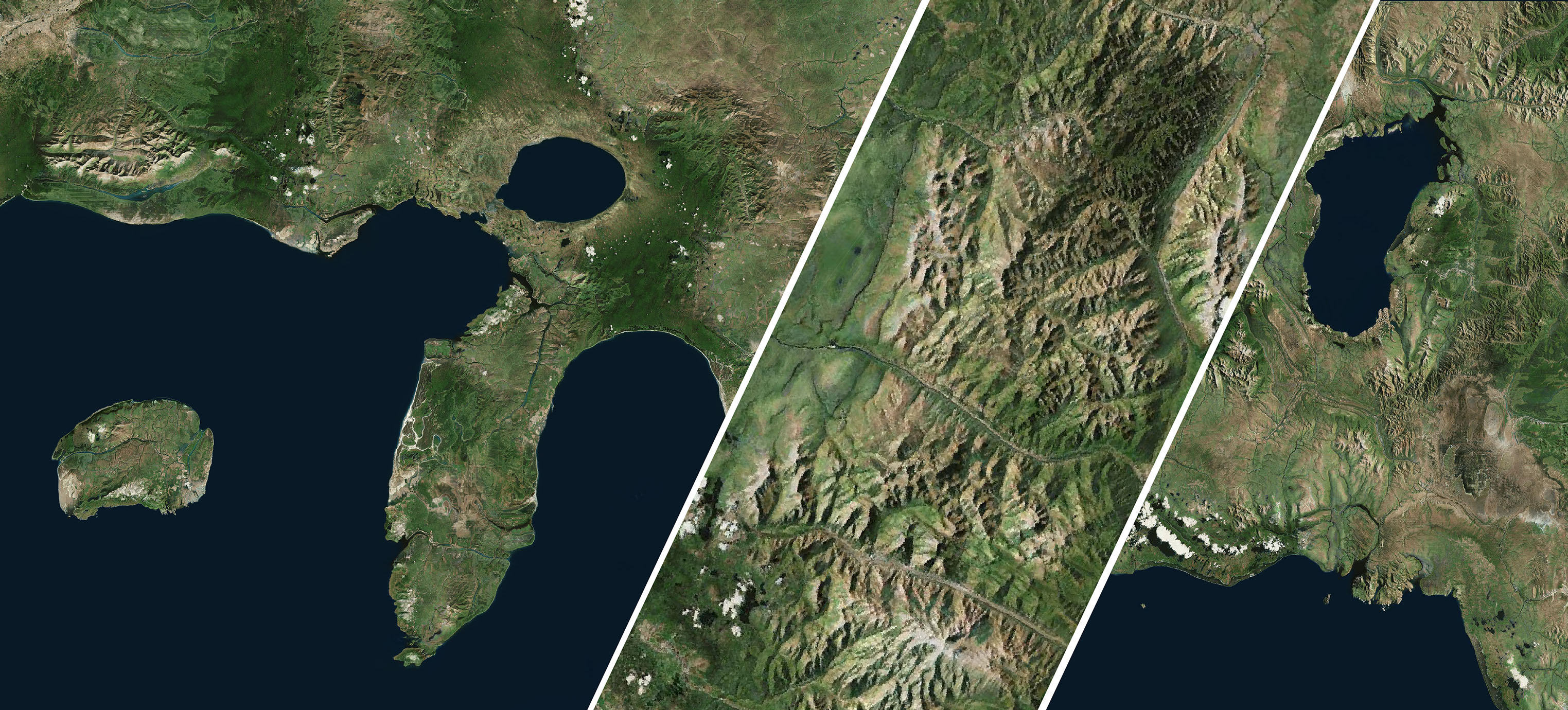}
  \caption{TileGAN can synthesize large-scale textures with rich details. We show aerial images at different levels of detail generated using our framework, which allows for interactive texture editing. Our results contain a broad diversity of features at multiple scales and can be several hundreds of megapixels in size.}
  \label{fig:teaser}
\end{teaserfigure}

\maketitle
\section{Introduction}
%
Example-based texture synthesis is the task of generating textures that look similar to a given input example. The visual features of the input texture should be faithfully reproduced while maintaining both small-scale as well as global characteristics of the exemplar.

In this paper, we are interested in synthesizing large-scale textures that consist of multiple megapixels (see Fig.~\ref{fig:teaser}). The first challenge in large-scale texture synthesis is to process a large amount of input data. This is crucial because without a considerable amount of reference data, any generated output will not have a lot of variability and lack features at multiple scales. Such a synthesized output could be large-scale, but will be very homogeneous and boring or repetitive. Recent work in parametric texture synthesis using generative adversarial networks (GANs) seems ideally suited to tackle this challenge and we build on a recent GAN architecture that can generate high-quality results when trained on natural textures~\cite{Karras2017ProGAN}. The second challenge in large-scale texture synthesis is how to generate large-scale output data. This is the core topic of this paper and we have identified two important sub-problems that we tackle in our work.

\begin{figure}[t]
  \centering
  \includegraphics[width=\linewidth]{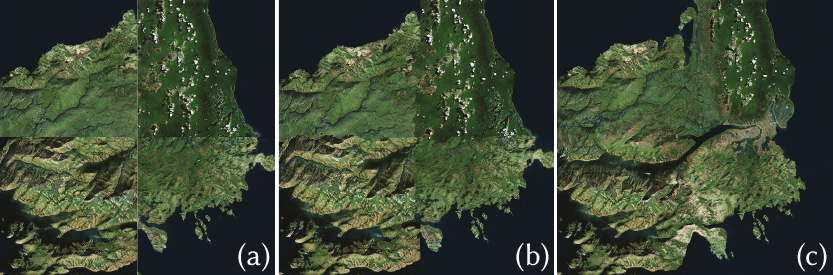}
  \caption{\bd{Tiling.} (a) Four input texture tiles. (b) Tiles are combined using graph cuts~\protect\cite{Kwatra2003Graphcut}. (c) Tiles are combined using our method.}
  \label{fig:problem}
\end{figure}

First, assuming that the selected GAN can only generate tiles of limited resolution, it is necessary to make these tiles match. There are multiple possible solutions to this problem that were explored in previous work. An elegant and powerful method is to compute graph cuts between overlapping tiles~\cite{Kwatra2003Graphcut}. While this method works well in some cases, very often it leads to artifacts when the blended tiles are not similar enough. Another possibility is to use a pixel-based texture synthesis algorithm like PatchMatch~\cite{Barnes2009PatchMatch} to repair seams between textures. A very simple method is to use blending. We propose a solution that is based on manipulating latent codes of lower resolution levels of the GAN to obtain nice transition regions. See Fig.~\ref{fig:problem} for a comparison of our method illustrated by blending four neighboring tiles.

Second, we need to be able to incorporate user input to control the visual appearance of the synthesized output. 
A major challenge for most existing texture synthesis methods is the artistic control over the final result. 
While patch-batch based texture synthesis techniques can be constructed to provide artistic control, such as painting by numbers~\cite{Hertzmann2001Image, Ritter2006Painting, Lukac15Brushables, Lockerman2016Multiscale}, most existing GAN-based texture synthesis approaches provide no (or minimal) artistic control. In this paper, we propose a solution based on latent brushes and other intuitive editing tools that allow easy global control on large scale texture maps. Please see the accompanying video for examples.

Technically, the major contribution of our paper is to provide a framework to take a GAN of limited resolution as a building block and produce a possibly infinite output texture. As a practical result, we are able to significantly improve the quality and speed of the state of the art in large-scale texture synthesis.

%

\section{Related Work}

We review the literature most related to our work sorted into multiple categories and refer the reader to Akl et al.~\shortcite{Akl2018STAR} for a more comprehensive and recent literature review on example-based texture synthesis.
\subsection{Non-parametric Texture Synthesis}
Existing non-parametric texture synthesis algorithms try to synthesize a new texture such that each $k \times k$ patch in the output texture has an approximate match in the input texture~\cite{Kwatra2005TOE}. A very important ingredient for these algorithms is a fast correspondence algorithm such as PatchMatch~\cite{Barnes2009PatchMatch} that is employed in most state-of-the-art texture synthesis algorithms, e.g.~\cite{Kaspar2015SSTO, Huang2015Single}.
PatchMatch can be extended to create faster queries~\cite{Barnes2015PatchTable} or additional error metrics~\cite{Darabi2012ImageMelding, Kaspar2015SSTO, Zhou2017Analysis}. 
While existing methods provide strong visual results, we propose to build on recent work in deep learning that shows a much greater promise with regards to the scalability of the considered input data or the size of the output due to faster synthesis speed. 
A notable earlier algorithm proposed a hierarchical extension using an earlier version of non-parametric synthesis~\cite{Han2008MTS}, but this algorithm is specific to the texture synthesis algorithm it employs~\cite{Lefebvre2005PCT} and it cannot be easily adapted to a deep learning framework.
\subsection{Parametric Texture Synthesis}
A popular early approach to texture synthesis was to extract features and feature statistics from an input texture and then try to create a new texture that would match these feature statistics~\cite{Heeger1995Pyramid, DeBonet1997Multiresolution, Portilla2000Parametric}. This idea has now been revisited using features extracted by neural networks. Gatys et al.~\shortcite{Gatys2015TextureSynthesis,Gatys2015Neural} proposed the idea to use inner products between feature layers at different levels of the network as a texture descriptor. For each layer of the network, each pair of features gives one inner product to compute. This idea was expanded by Sendik and Cohen-Or~\shortcite{Sendik2017DCor}, who introduced a structural energy term based on correlations between deep features, thus capturing self-similarities and regularities in the structural composition of the texture. The technique proposed by Snelgrove et al.~\shortcite{Snelgrove2017High} presents an early effort to increase the maximum size of texture features that can be synthesized using the method of Gatys et al.~\shortcite{Gatys2015TextureSynthesis}. This is accomplished by matching a small number of network layers across many scales of a Gaussian pyramid leading to improved synthesized textures. Instead of using gradient-based optimization to compute new textures, it is also possible to train a generator using the feature statistics for the loss function~\cite{Dosovitskiy2016Generating}. 

\begin{figure*}[t]
  \includegraphics[width=\textwidth]{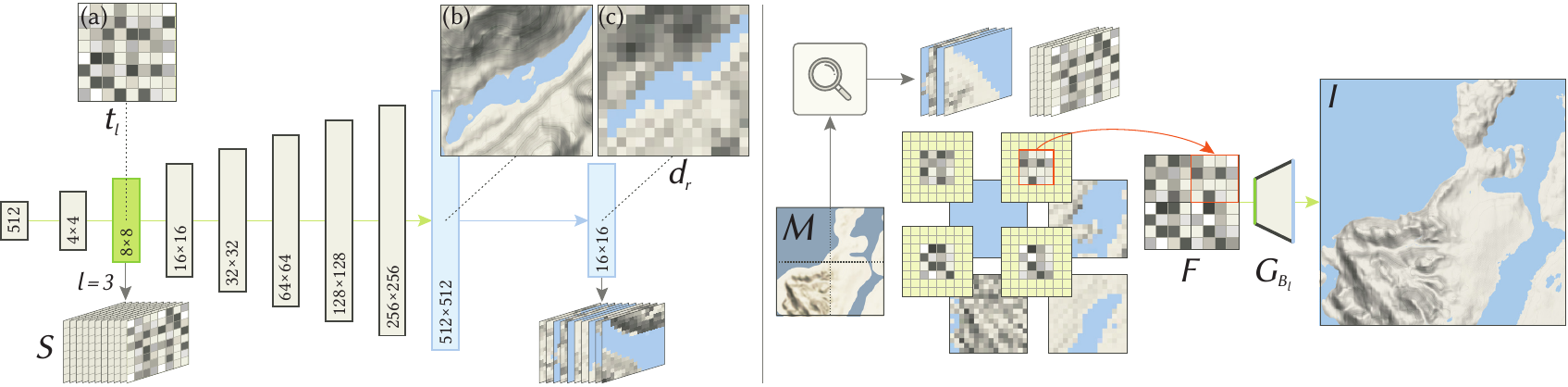}
	\caption{{\bf Synthesis overview.} Our generator network (left side) takes a random latent vector of size 512 as input. At latent level $l=3$, the latent tile (a) is of spatial resolution $8\times8$. For simplicity, the depth of the feature maps in each layer is not shown. We sample a large set $S$ of latent tiles for processing. For each sampled latent tile we also generate the corresponding output image (b) and save a downsampled version $d_r$ (c) (here we set $r=16$) for retrieval during neighborhood similarity matching. During synthesis (right side), we look up tiles according to the similarity of their downsampled representation $d_r$ to the corresponding region in the guidance image $M$ and tile them to generate the latent field $F$. The size of the output and the amount of blending can be adjusted by cropping each latent tile when generating the field. Finally, $F$ is processed by $G_{B_l}$ to produce the final output image $I$.
	}
	\label{fig:overview}
\end{figure*}

\subsection{GANs Trained for Texture Synthesis}
Generative adversarial networks (GANs) were introduced in a seminal paper by Goodfellow et al.~\shortcite{Goodfellow2014GANs}. Over the years, the architecture of GANs has improved significantly and state-of-the-art GANs can produce results of stunning visual quality~\cite{Karras2017ProGAN, Brock2018BigGAN, Karras2018StyleGAN, Zhang2018SAG}.
In our work, we have chosen to build on the framework proposed by Karras et al.~\shortcite{Karras2017ProGAN}.
They introduced a progressively growing architecture that starts the training on a low-resolution exemplar and slowly increases the size of the networks, as well as the exemplars. Their network is able to produce semantically coherent image content at a significantly higher resolution than previous work. 
Zhou et al.~\shortcite{Zhou2018TexSynth} introduced a technique to expand textures while preserving challenging structural arrangements by iteratively training a GAN on sub-blocks of the input textures. While this work uses GANs, it only uses a single image as input by generating many different crops during training.
%
%
The convolutional nature of GANs can be exploited to synthesize images and textures of output sizes different from the image resolution the GAN was trained on \cite{Jetchev2016SGAN, Bergmann2017PSGAN}. GANosaic~\cite{Jetchev2017GANosaic} extends such methods to generate textures by optimizing the latent noise space to produce textures that match the overall content of a given guidance image. However, such methods are limited in the type of textures they support, the expected size, and the overall variability and quality of the output.
The image stylization method FAMOS~\cite{Jetchev2018FAMOS} improves on the quality by training the texture GAN and the guidance image styling network at the same time. While this method produces smoother transitions between texture patches, it still suffers from issues relating to scalability and variability, which we try to address in our method.
\subsection{Selected Applications of GANs}
GANs have been successfully adapted to classical image processing problems, such as inpainting~\cite{Pathak2016ContextEncoders, Yeh2017SemanticInpainting, Yu2018GenerativeInpainting} and super-resolution~\cite{Wang2018ESRGAN}. 
While traditional GANs generate images starting from a random vector, the GAN training can be extended to the problem of image-to-image translation using either paired or unpaired training data~\cite{Isola2016Pix2Pix, Zhu2017CycleGAN, Zhu2017Towards, Huang2018MUNIT}.
In computer graphics, recent papers apply GANs to the synthesis of caricatures of human faces~\cite{Cao2018Cari}, the synthesis of human avatars from a single image~\cite{Nagano2018paGAN}, texture and geometry synthesis of building details~\cite{Kelly2018FrankenGAN}, surface-based modeling of shapes ~\cite{Ben-Hamu2018MGS} and the volumetric modeling of shapes~\cite{Wang2018Global}.
The most related problem to our work is the problem of terrain synthesis~\cite{Guerin2017IET}. 

\section{Overview}
In this section, we present the three main components of our framework in a high-level overview:

\paragraph{Generative models.} Our framework requires a generative model that produces novel images. State-of-the-art GANs typically consist of two networks: a generator and a discriminator. The generator network produces sample images which match the training distribution using convolutional layers that gradually increase the spatial resolution of a random latent vector to a full-size image. The discriminator network assesses how well the generated samples match the training distribution. The two networks are constructed to be differentiable and their gradients are used to guide the training of the full generative model. Our main focus in this paper is on combining multiple outputs of the generator network of a standard GAN for large scale texture synthesis. More details on the GANs and data sets we use in our experiments are given in Sec.~\ref{sec:implementation}.

\paragraph{Synthesis.} Our key contribution is a method to synthesize plausible large-scale non-homogeneous textures using a pretrained generator network. This is accomplished by generating a tiling of compatible intermediate latent vectors, which we call the \emph{latent field} $F$, that the generator network $G$ uses to produce a coherent large-scale texture $I$ (see Fig.~\ref{fig:overview}). The intermediate latent vectors can be efficiently sampled and stored for analysis and online processing. In order to ensure that the synthesized textures are globally coherent, we optimize the latent field to satisfy two main objectives. First, the expected synthesized output should follow an initial small target guidance map $M$ for the expected large-scale synthesized image. This map can be randomly generated or specified by the user. Second, in order to afford local coherence and minimize abrupt texture changes between neighboring texture tiles, we optimize the latent field by replacing problematic tiles with better candidates that are more compatible with their neighbors. The details of our entire synthesis pipeline will be presented in Sec.~\ref{sec:synthesis}.

\paragraph{Artistic Control.} We propose a set of tools to facilitate user control over our texture synthesis process. The key idea behind the control our method affords during synthesis lies in modifying the latent field. To that end, we utilize operations such as painting, shuffling, copying, and target image matching, all of which enable different ways of artistic control.
We will discuss details about our interactive tool in Sec.~\ref{sec:control}. 

\section{Synthesis}
\label{sec:synthesis}

We first describe the general notation used in this paper before describing the different phases of our framework. We start by redefining the generator network, from a standard deep convolutional GAN, as:
\begin{equation}
G(z)=G_{B_l}(G_{A_l}(z)),
\end{equation}
where $z$ is a randomly sampled latent vector and $l$ specifies the intermediate level at which we plan to perform our latent field synthesis. Lastly, $G_{B_l}$ and $G_{A_l}$ are two parts of $G$ that split the set of convolutional layers at the level $l$ (see Fig.~\ref{fig:overview}). For a GAN with $n$ levels, $G_{A_l}$ takes the latent vector at level 1 as input and produces a $k \times k$ tile at level $l$ and $G_{B_l}$ takes a tile at level $l$ as input and produces a color image at level $n$, the final level.

Our large-scale non-homogeneous texture synthesis framework is divided into three phases: (1) a one-time preprocessing phase, (2) an online latent field synthesis phase, and (3) an online texture synthesis phase.

\subsection{Preprocessing}
\label{sec:preprocessing}
The first step of preprocessing is to create a large set of texture samples $S$ that are generated using the generator network from a standard deep convolutional GAN. Each sample $s_i$ comprises two components: (1) an intermediate tensor $t_l = G_{A_l}(z)$, which we refer to as a \emph{latent tile} where $l$ is the level at which we will synthesize the latent field, and (2) $d_r$ a downsampled version of the texture map $G_{B_l}(t_l)$, where $r$ represents its spatial resolution. The greater the number of samples in $S$, the more texture variability is afforded by our framework. The second step in this phase is to cluster the texture samples in $S$ by their visual appearance, using $d_r$, in order to enable fast lookup of visually similar latent tiles.
We perform the clustering using $k$-means and assign cluster centers $c_k$ as representative texture samples.

\begin{algorithm}[t]
\caption{\sffamily{The TileGAN Algorithm}}
\label{alg:tilegan}
\begin{algorithmic}[1]
\Procedure{GenerateTextureMap}{}\newline
\textbf{input:} $G_{B_l}, S, M$\newline
\textbf{output:} $F$, $I$
\State{$U \gets \textsc{NextUnassignedPatch}~(F)$  }
\While{$\textsc{Count}~(U) > 0 $} \Comment{\sffamily{\small{Initial tiling}}}
  \State{{$i \gets U(0)$}}
  \State{{$F_i \gets \textsc{TopMatch}~(i, F, I, S, M)$ \Comment{\sffamily{\small{Single top match}}} }}
  \State{{$I_i \gets G_{B_l}(F_i)$}}
\EndWhile
\While{$E(F) > \theta $} \Comment{\sffamily{\small{Optimizing the entire texture}}}
  \State{{$i \gets \textsc{Random}~(F, I)$}}
  \State{{$F_i \gets \textsc{BetterMatch}~(i, F, I, S, M)$}}
  \State{{$I_i \gets G_{B_l}(F_i)$}}
\EndWhile
\State{\Return $F$, $I$}
\EndProcedure
\end{algorithmic}
\end{algorithm}

\begin{algorithm}
\caption{\sffamily{Refine Latent Tile}}
\label{alg:bettermatch}
\begin{algorithmic}[1]
\Procedure{BetterMatch}{}\newline
\textbf{input:} $i, F, I, S, M$ \Comment{\sffamily{\small{Relevant neighboring regions}}}\newline
\textbf{output:} $F_i$
\If{$E(F_i) \leq \theta $}
    \State{\Return $F_i$}
\EndIf
\State{{$T \gets \textsc{TopMatches}~(i, F, I, S, M)$} \Comment{\sffamily{\small{Set of top matches}}}}
\State{\Return $\argmin_{i} E(T_i)$ }
\EndProcedure
\end{algorithmic}
\end{algorithm}

\subsection{Latent Field Synthesis}
\label{sec:latent-field-synthesis}

\paragraph{Markov Random Fields.} The second phase of our framework is the synthesis of large compositions of GAN-generated textures with no apparent visual artifacts, seams, or obvious repetition. We use a variant of the Markov Random Fields (MRF) model for texture synthesis applied on the latent field $F$. While the MRF model has been applied to texture colors~\cite{Wei2009STAR} as well as texture statistics~\cite{Li2016MRF}, we are the first to propose an application to GAN latent vectors. The goal of such an MRF model can be redefined for our framework as follows: given a large set of individual texture tiles sampled from a single distribution, synthesize a large scale output of texture tiles so that for each output tile, its spatial neighborhood is similar to some neighborhood from the input distribution. With this MRF assumption, the similarity of the local neighborhood between input and output help ensure an overall coherent texture map with minimal boundary artifacts. 

\paragraph{Two steps process.} In order to efficiently generate textures at large scale, we perform the latent field synthesis in two steps: an initialization step and an iterative refinement step. Alg.~\ref{alg:tilegan} formalizes the entire process of our framework for latent field synthesis. Splitting the computational task of synthesizing the texture facilitates interactive editing. The first step is typically computed on the order of seconds and is immediately presented to the user. The refinement step is computed on a background process that regularly updates the latent field and displays the final output. Alg.~\ref{alg:bettermatch} represents how we find better candidates in the refinement step.

\subsubsection{Initialization} 
In the initializing step, we aim to efficiently generate a tiling of the latent field that approximately satisfies the guidance map $M$. The map $M$ provides global content control. At this stage, we assume that the texture samples $S$, generated at the latent level $l$, and its clustering result were computed in a prior preprocessing step. We start by performing a latent-tile-based texture synthesis to cover all unassigned tiles in the output latent field $F$. For each unassigned tile, we find the single top matching $F_i$ using the unary energy term defined below. We repeat this processing until no unassigned latents remain.

\begin{figure}[t]
	\centering
	\includegraphics[width=\linewidth]{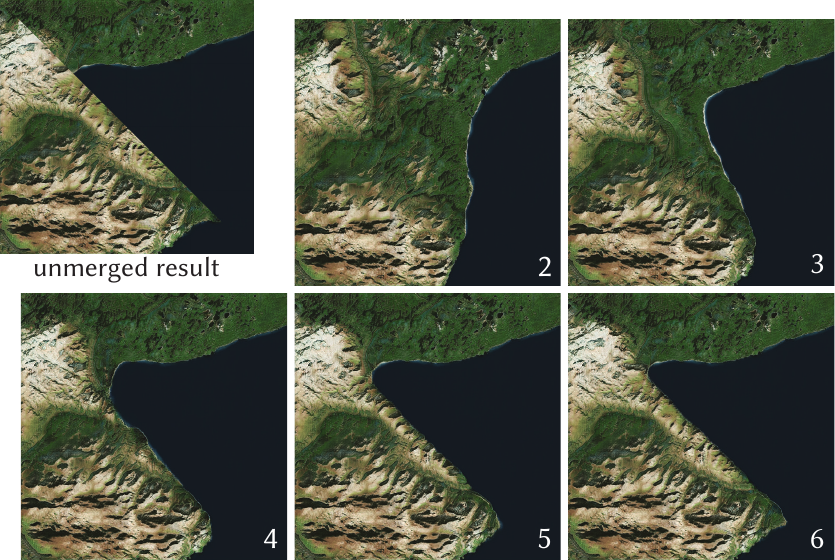}
	\caption{\bd{Latent merging.} Our network architecture creates transitions between very different tiles when merging latent vectors at various depths. Here we show merging results for layers two to six. The impact region of the transition decreases in size when processing the latents at later stages in the network.}
	\label{fig:merging}
\end{figure}

\subsubsection{Refinement}
\label{sub:Refinement}

Refinement steps are performed until the total latent field's energy is lower than our set threshold. This stopping criterion is currently set empirically to a value that ensures that a desirable variety of visual features in the tiling is preserved during the MRF refinement. In each step, we randomly sample a latent tile in $F$ and check for candidate tiles that minimize the local energy. We define the optimal latent field as the field that minimizes the following energy of weighted unary and binary terms:
\begin{equation}
E = E_{m} + E_{n}.
\end{equation}
The unary term $E_{m}$ is the sum of visual similarities of $d_r$ of a candidate tile $F_i$ with its corresponding region in $M$. In our experiments, we consider the Euclidean distance of the two images as the similarity measure in order to accelerate this computation. The binary term $E_{n}$ considers the 4-connected neighboring latent tiles for each tile by the following weighted dissimilarity terms:
\begin{equation}
E_{n} = \frac{1}{2} \sum_{i}\sum_{j \in N_i}(\lambda_V D_{V}(F_i,F_j) + \lambda_L D_{L}(F_i, F_j) + \lambda_C D_{C}(F_i,F_j)).
\end{equation}
These terms represent the dissimilarity between a tile $F_i$ and another tile in the set $T_i$ of its 4-connected neighbors: visual appearance $D_{V}$, latent vector representation $D_{L}$, and cluster membership $D_{C}$. For every pair of latent tiles in the 4-connected neighborhood, we approximate $D_{V}$ and $D_{L}$ using the Euclidean distance of their overlapping region. The dissimilarity measure $D_{V}$ is computed on the corresponding $d_r$ of each tile while $D_{L}$ is computed on the corresponding latent tensor $t_l$. The last term $D_{C}$ is the average agreement of cluster membership where we assign a 0 to pairs with matching clusters and 1 to non-matching pairs. The different energy terms are weighted by the corresponding $\lambda_x$ weight parameter. We set $\lambda_V$ as 1 and $\lambda_L$ and $\lambda_C$ as 0.5 in our experiments. When finding the top matches in the refinement step, we first return the top $10$ matching tiles using $E_{m}$ and then compute the entire energy after placing each candidate tile. 

\paragraph{Boundary and latents tiles.} An important aspect when combining tiles from different samples of $S$ is the unpredictability at their joining region (see Fig.~\ref{fig:merge_failure}). In our experiments, we have noticed that latents that fall on the outer most regions of the latent tile exhibit lots of instability. This is likely due to a bias caused by the zero padding that is applied in $G_{B_l}$. In order to minimize this effect, we only consider a cropped version for each sample of $S$. The size of the cropped latent tile influences the overall visual coherence of neighboring output regions, where smaller latent tiles exhibit a smoother feature blending than larger tiles. In our experiments, we have typically used latent tile sizes of $2\times2$ to $4\times4$ regardless of the merge level $l$. While we crop the latent blocks, we use the entire representative image $d_r$ for comparison with the guidance map, which creates an overlapping sliding-window effect when finding tile matches, thereby further enhancing the coherence of neighboring tiles.

\paragraph{Choice of parameter $l$.}
Selecting the level at which to split the GAN is a trade-off between the quality of the transition region and the scope of the region impacted by the transition changes (see Fig.~\ref{fig:merging}). We have mainly experimented with splitting at earlier levels $l=2$ to $l=5$ in our work because these parameters yield the best visual results according to our judgment.

\begin{table*}[t]
\sffamily
\centering
\begin{tabularx}{\textwidth}{r|llcc|lll}
\toprule
\bd{Result}     & \bd{Training data set}    & \bd{Output}       & \bd{Merge level}  & \bd{Latent tile size} &\bd{Synthesis} &\bd{User editing}  &\bd{Total}  \\ 
                & \sm{(Megapixels)}         & \sm{(Megapixels)} &                   &                       & \sm{(min)}    & \sm{(min)}        & \sm{(min)} \\ 
\midrule
Medieval Island    \sm{(Fig.~\ref{fig:results1} top)}       &  4700 & 94   &    2 & $2~\times~2$ & 15.0  &   1  &  16.0 \\ 
Dinosaur Park      \sm{(Fig.~\ref{fig:results1} bottom)}    & 17000 & 22   &    2 & $2~\times~2$ & 4.5   &   5  &   9.5 \\ 
Mountain Painting  \sm{(Fig.~\ref{fig:results2} top)}       &  7800 & 620  &    2 & $2~\times~2$ & 25.0  &   0  &  25.0 \\ 
Space Panorama     \sm{(Fig.~\ref{fig:results2} bottom)}    &  5000 & 6.5  &    5 & $1~\times~1$ & 1.2   &   10 &  11.2 \\ 
\midrule
STTO               \sm{(Fig.~\ref{Fig:comparison} top)}     & 0.27  & 1    &  --- &          --- & 9.4   &  --- &   9.4 \\
NSTS               \sm{(Fig.~\ref{Fig:comparison} top)}     & 0.27  & 1.08 &  --- &          --- & 2000  &  --- &  2000 \\
TileGAN            \sm{(Fig.~\ref{Fig:comparison} top)}     & 17000 & 9.8  &    2 & $2~\times~2$ & 3.0   &    0 &     3 \\
\midrule
STTO               \sm{(Fig.~\ref{Fig:comparison} bottom)}  & 0.27  & 1    &  --- &          --- & 10.25 &  --- & 10.25 \\
NSTS               \sm{(Fig.~\ref{Fig:comparison} bottom)}  & 0.27  & 1.08 &  --- &          --- & 2000  &  --- &  2000 \\
TileGAN            \sm{(Fig.~\ref{Fig:comparison} bottom)}  & 17000 & 9.8  &    2 & $2~\times~2$ & 3.0   &    0 &     3 \\
\bottomrule
\end{tabularx}
\bigskip
\caption{\textbf{Quantitative evaluation.} At the top of the table, we show an evaluation of results of our system presented in Figs~\ref{fig:results1} and~\ref{fig:results2}. Below is a comparison between results generated using TileGAN and the state-of-the-art techniques of Self-Tuning Texture Optimization and Non-Stationary Texture Synthesis.}
\label{tab:quantresults}
\end{table*}

\subsection{Texture Synthesis}
\label{sub:texture-synthesis}

The final synthesized image is generated by taking a latent field and applying the trained generator network $G_{B_l}$. Using this multi-stage process, the network is able to output arbitrarily large results.
A latent field of size $w \times h$ will result in an RGB texture of size $2^{(n-l)}w \times 2^{(n-l)}h$, where $n$ is the number of levels in the pyramid.
We use $l=3$, $n=9$ for most of our experiments. Since the generating function $G_{B_l}$ is convolutional, we can profit in two ways. First, it can be efficiently applied for local re-synthesis. Second, arbitrarily large latent fields can be processed by multiple overlapping applications of $G_{B_l}$ where the overlapping parts of the output of each application of $G_{B_l}$ is discarded.

\begin{figure}[t]
	\centering
	\includegraphics[width=\linewidth]{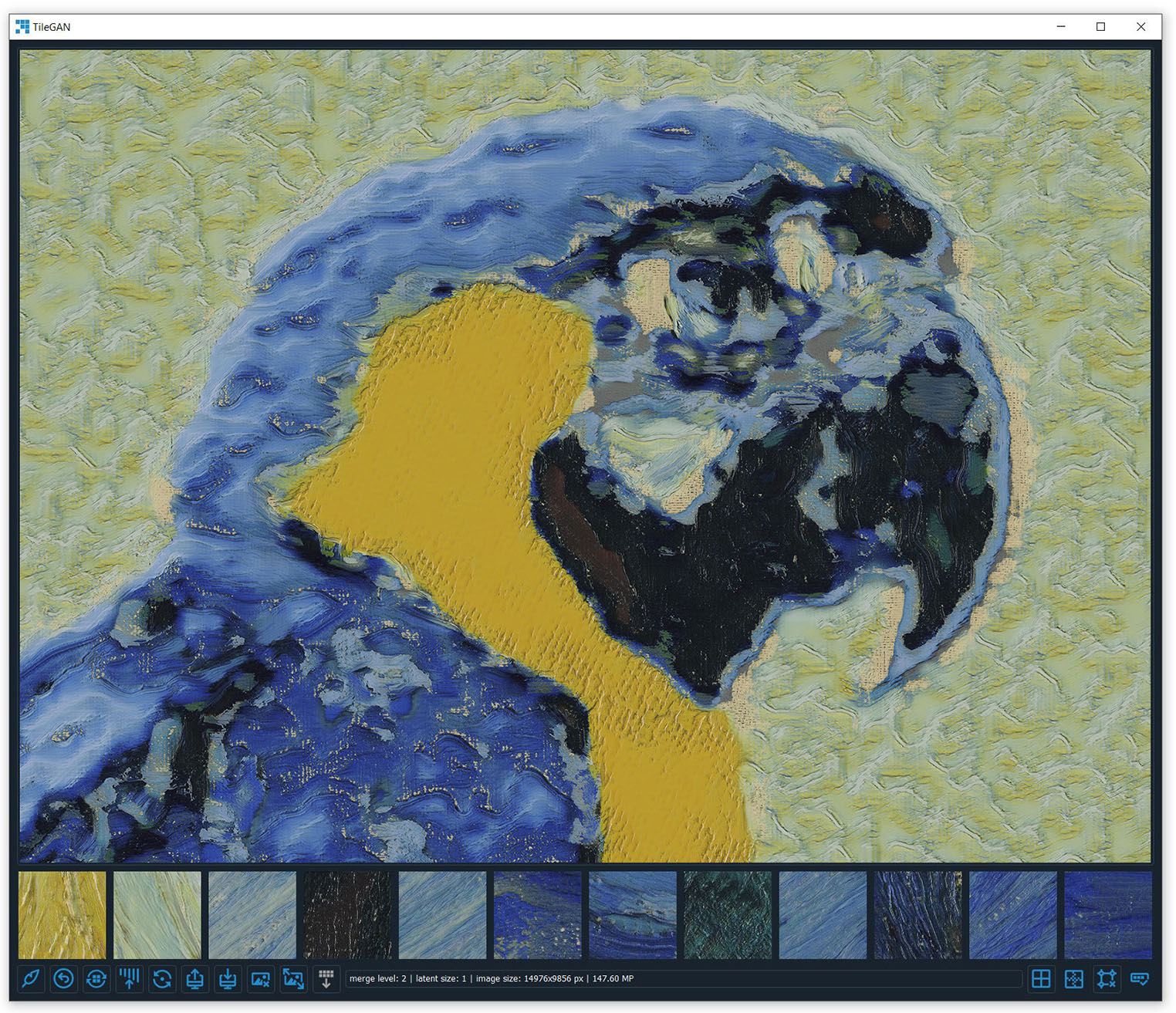}
	\caption{\textbf{Artistic control.} Our interface allows users to manipulate latent fields interactively. The user can select images as guidance maps and edit the image using various latent-space painting techniques. On the bottom of the interface, representative tiles show clustered latents, which can be used to manipulate the image content. The main part of the UI consists of a preview of the final texture that is locally updated as soon as changes to the latent field are made. More details on user interaction in Section~\ref{sec:control}.}
	\label{fig:ui}
\end{figure}

\section{Artistic Control}
\label{sec:control}

Our texture synthesis framework can fully automatically generate plausible results. However, as with most texture synthesis scenarios, user control and interactive editing are highly desirable. We have developed an interactive tool with different editing operations for our GAN-based image synthesis approach, see Fig.~\ref{fig:ui}. 

We provide two major sets of editing operations: (1) directly manipulating the latent field, (2) editing the guidance map. 

The first editing operation for manipulating the latent field allows the user to drag and drop a tile from a list of clusters (Fig.~\ref{fig:ui}, bottom) onto an existing latent field. The inserted latent tile is randomly sampled from the latent tiles belonging to the selected cluster. In order to visualize the expected tile appearance, we show the user a representative image corresponding to the cluster centers $c_k$. This tool can be generalized as a GAN-based paintbrush of variable size, which offers a high degree of user control. The second editing operation is a cloning tool, where we take parts of existing content from the synthesized image and clone the respective latents onto other regions. We provide an option to spatially shuffle the cloned tiles to add more diversity to the cloned region. Moreover, we can add small amounts of noise or interpolate between two latent tiles to allow for additional degrees of variability. These simple latent manipulation tools provide local control of the resulting output. 

Finally, the appearance of the output texture can be influenced by modifying the guidance map using traditional image manipulation techniques. This capability provides virtually limitless variations in the size of the output, shading, placement of features, etc.

\section{Implementation Details}
\label{sec:implementation}

\paragraph{Learning.}
Our GAN architecture and training is based on the approach of Karras et al.~\shortcite{Karras2017ProGAN}, called Progressive Growing of GANs (ProGAN). We have slightly modified the generator architecture to extract the intermediate latents at any arbitrary layer of the GAN. These latent tiles can be modified and merged to generate large-scale non-homogeneous textures. We have chosen ProGAN over other GAN architectures because it consistently produces high-quality output images and because the architecture consisting of a stack of identical building blocks facilitates the division into the two parts that allow us to manipulate the intermediate latent field.

While the training process may take on the order of days training on multi-GPUs to reach tiles of acceptable quality, the synthesis and editing steps of the texture generation are possible at interactive rates running on a machine with a single GPU. The preprocessing step is done only once and typically requires 30 minutes to sample a set $S$ of size $100K$ latent tiles and then cluster the corresponding representative $d_r$ images into $k=10$ clusters.
For each data set, we train the GAN on four NVIDIA v100 GPUs for $16K$ iterations for around 4 days. We use the default optimizer and training schedule provided by the official TensorFlow~\cite{Tensorflow2015} implementation from \cite{Karras2017ProGAN}\footnote{\label{ref:progan}\href{https://github.com/tkarras/progressive_growing_of_gans}{\texttt{github.com/tkarras/progressive\_growing\_of\_gans}}}.
%
 
\paragraph{Training Data.}
We have compiled various training data sets for experimentation with our method. Popular existing data sets like CelebA or LSUN are not suitable for large-scale texture synthesis.
Therefore, we have curated our own test data from several sources of publicly available large-scale imagery, all of which are processed as image tiles of size $512 \times 512$:
\begin{itemize}
    \item {\bf Terrain map.} We collect $18K$ tiles of the terrain basemap provided by Google Maps.
    \item {\bf Satellite imagery.} We use $65K$ samples from the tiles of Landsat satellite images.
    \item {\bf Oil canvas.} We also consider high-resolution images of smaller objects including $30K$ tiles from the detailed Gigapixel image of the Vincent van Gogh's The Starry Night provided by the Google Art Project.
    \item {\bf Night sky.} We sample a total of $19K$ high-resolution tiles from the European Southern Observatory and from the Hubble Space Telescope image repository.
\end{itemize}
For all data sets, we do not perform any alignment steps or augmentation of the input training tiles.

\begin{figure*}[t]
  \includegraphics[width=\textwidth]{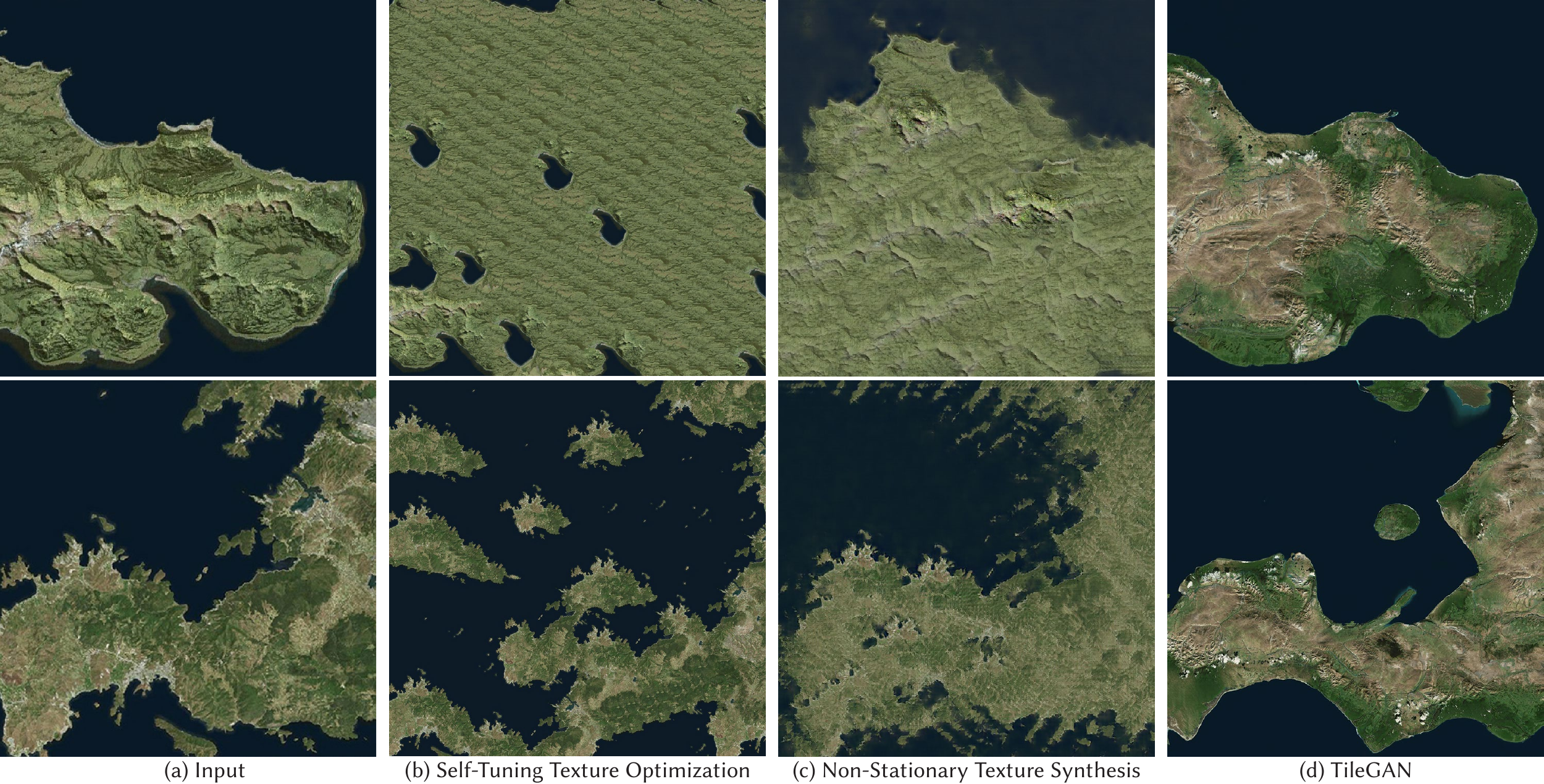}
  \caption{\bd{Comparison.} We provide an informal comparison to the current state-of-the-art texture synthesis methods Self-Tuning Texture Optimization (b) and Non-Stationary Texture Synthesis (c), using (a) as input. We can observe that the current state of the art has difficulty in reproducing multi-scale features while our result (d) exibits interesting variations. Note that our method does not use (a) as input, but only as guidance for synthesis based on pre-trained content.}
  \label{Fig:comparison}
\end{figure*}

\section{Results}
\label{sec:Results}

In this section, we present a qualitative and quantitative analysis of the results created using our method. All synthesis results are generated using our interactive tool written in Python and running on a desktop machine equipped with an Intel Xeon 3.00GHz CPU with 32GB RAM and a single NVIDIA TITAN Xp GPU with 12GB memory. Note that in order to handle results with hundreds of megapixels, we resort to the caching of the latent field and synthesizing the image in chunks by the maximum supported block that fits on the GPU one block at a time.

\begin{figure*}[t]
	\centering
	\includegraphics[width=\linewidth]{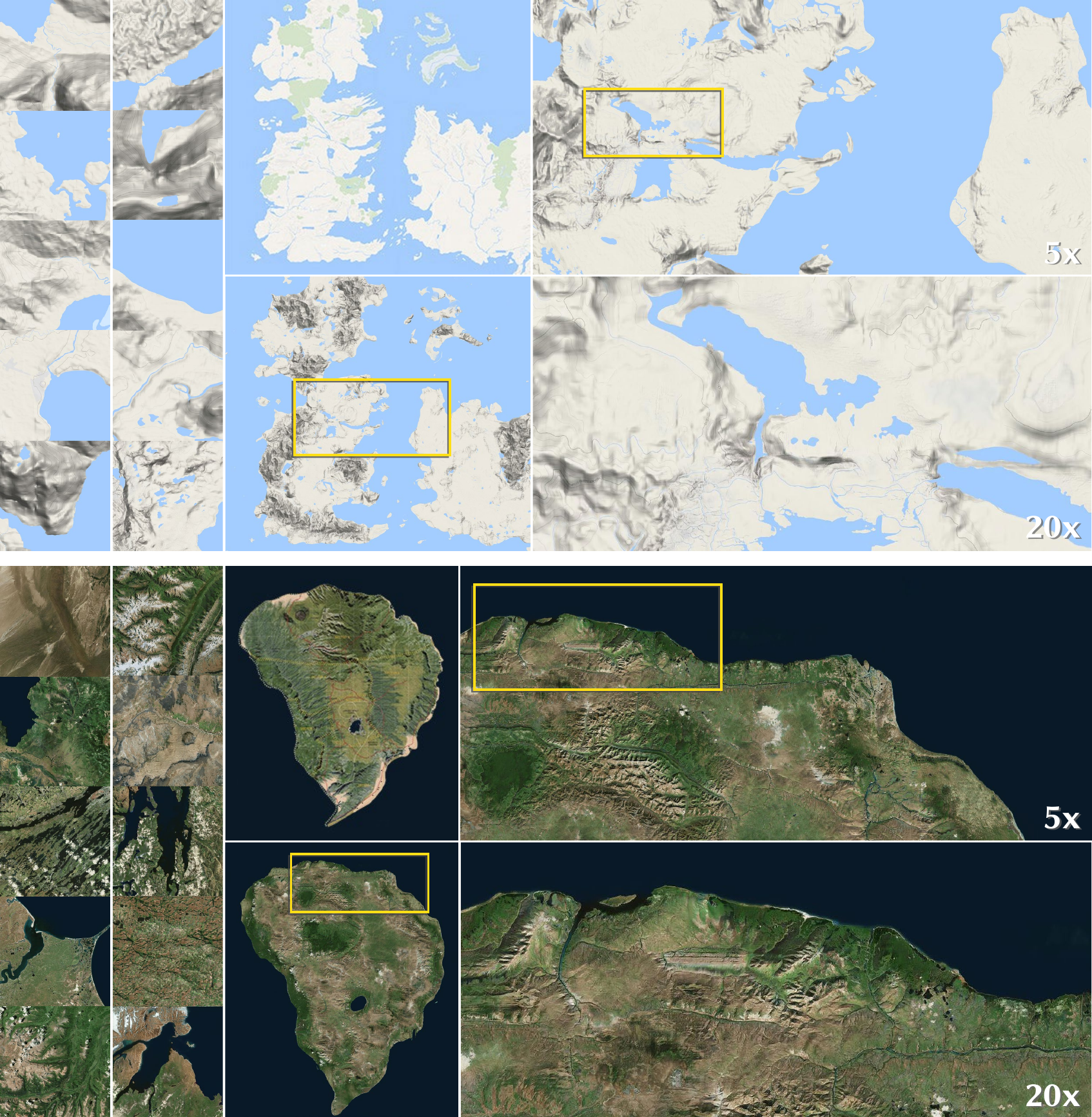}
	\caption{
	    \textbf{Results generated using TileGAN.} The results were generated using GANs trained on the Google terrain map (top) and the Satellite imagery (bottom) data sets. For each result, the first column shows samples from the training data, the second column contains sample tiles generated by the trained GAN, the third column features the low-resolution guidance map input to our method and the final large-scale output. In the last column, we show two cropped and zoomed regions scaled by the zoom value given on the bottom right. \imagecredits{top, first column \textcopyright~Google; bottom, first column \textcopyright~ESRI.}
	}
	\label{fig:results1}
\end{figure*}

\begin{figure*}[t]
	\centering
	\includegraphics[width=\linewidth]{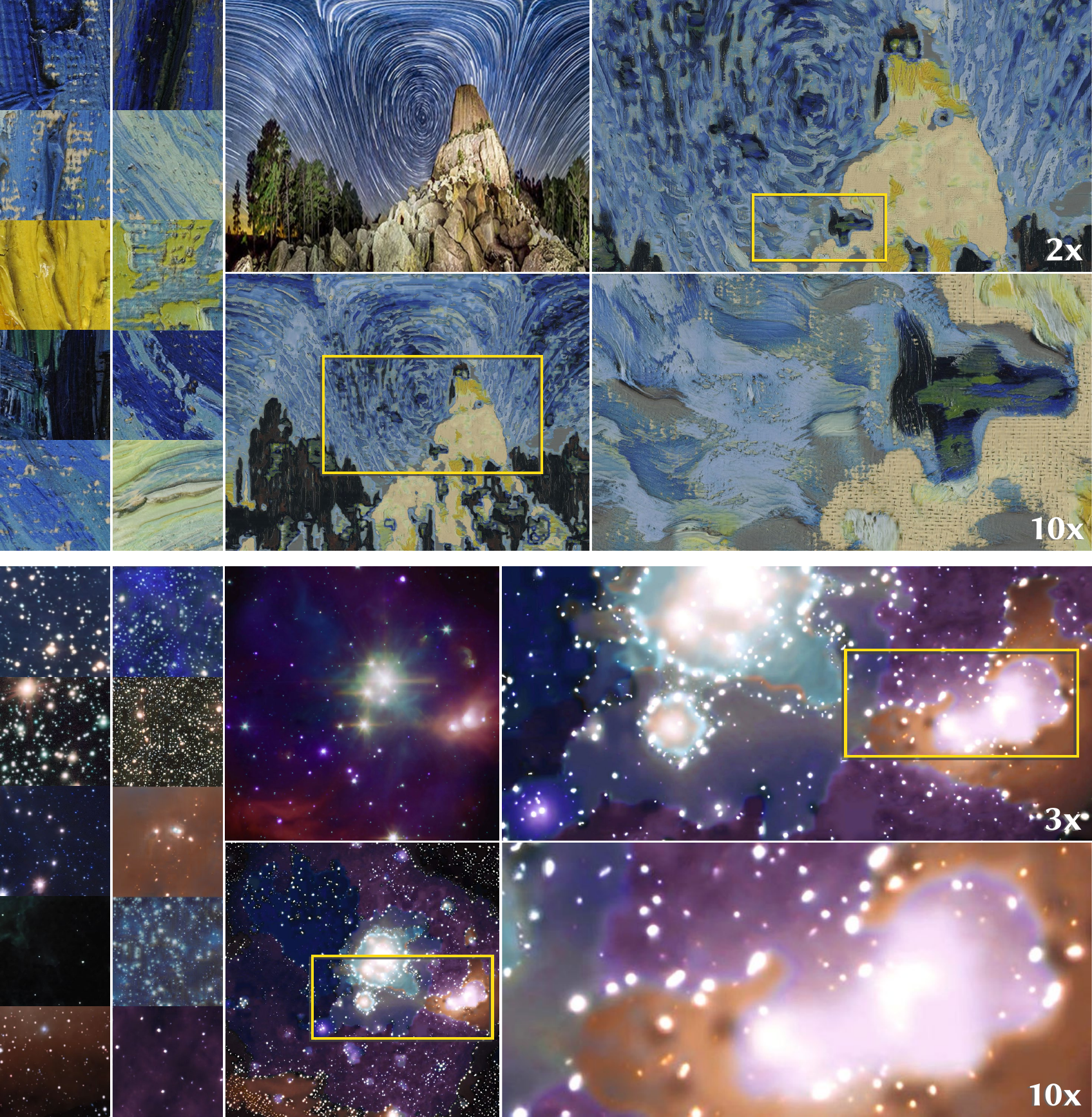}
	\caption{
	   \textbf{Further TileGAN results.} These results were generated by our method using GANs trained on the Oil canvas (top) and the Night sky (bottom) data sets. See Fig.~\protect\ref{fig:results1} for an explanation of the image columns. \imagecredits{top, low-resolution input \textcopyright~Vincent Brady; bottom, first column \textcopyright~ESO and ESA/Hubble.}
	}
	\label{fig:results2}
\end{figure*}

\paragraph{Visual quality.}
We use our framework on the data sets described in Sec.~\ref{sec:implementation} and showcase selected results in Fig.~\ref{fig:results1} and Fig.~\ref{fig:results2} to demonstrate the quality and variability afforded by our method. Tab.~\ref{tab:quantresults} shows the corresponding statistics for each result. We argue that our method can generate large-scale non-homogeneous textures with high visual quality that surpasses the quality achieved by any other published method.

\paragraph{Visual comparison to the state of the art.} 
We compare our method to two other state-of-the-art algorithms. We selected self-tuning texture optimization (STTO)~\cite{Kaspar2015SSTO}, which we believe to be the state-of-the-art texture synthesis algorithm not using neural networks and non-stationary texture synthesis (NSTS)~\cite{Zhou2018TexSynth}, a recent neural network-based algorithm.
We take a single texture tile of resolution $628 \times 425$ as input for the STTO and NSTS algorithms. For both techniques, we used the recommended settings provided by the authors. To generate our results, we use the fully trained network and apply the input image as a guidance map. We present a comparison of the different methods in Fig.~\ref{Fig:comparison}.
Even though we could verify that STTO generates excellent results for a large variety of textures, in these tests we can see that STTO is unable to handle the multi-scale features present in the aerial image and produces a highly repetitive output. This demonstrates that challenging issues related to multi-scale texture synthesis have not been fully explored.
We also verified our installation of the NSTS code released by the authors by replicating the results in their paper before running it on aerial images. However, NSTS is also not able to produce high-quality results when taking aerial images as input.
We believe that multi-scale texture synthesis requires a large amount of input and that previous work is inherently not suitable to generate multi-scale content of high quality. By contrast, we can generate images of high visual quality with few artifacts.

\paragraph{Quantitative comparison.} 
We also provide a comparison of the running time and scale for the synthesis of various examples and methods (see Table~\ref{tab:quantresults}). As shown in the table, self-tuning texture optimization does not scale as well with respect to the input data size and the output data size as our method. 
Non-stationary texture synthesis spends a lot of time on analyzing a small input texture, but this is not a suitable strategy for large-scale and multi-scale texture synthesis. Our method is faster than the competing methods if we exclude our preprocessing times with the justification that the preprocessing time would be amortized over many applications of a single trained GAN generator.

\begin{figure}[t]
	\centering
	\includegraphics[width=\linewidth]{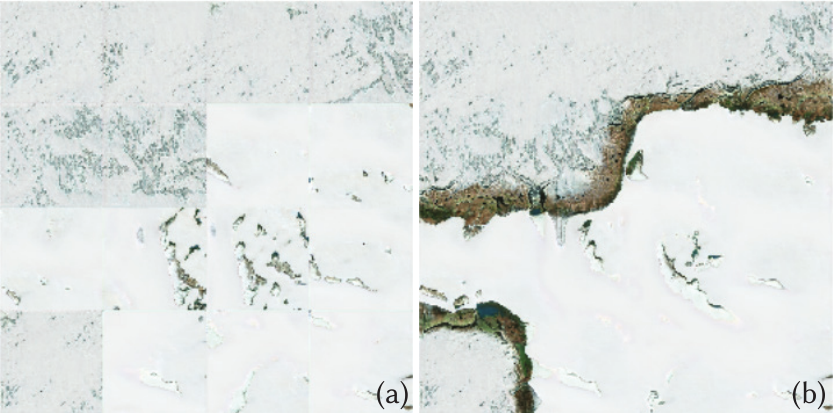}
	\caption{\bd{Undesired transitions.} When tiling latents with apparently similar visual appearance, such as this extreme case depicting snow and ice (a), our framework can produce unexpected regions of very different appearance (b). While these transitions look plausible, they can introduce unintended salient boundaries. We believe that this is due to the fact that the ice samples were learned from different continents and that no transition between these regions is available in the training data.}
    \label{fig:merge_failure}
\end{figure}

\section{Limitations}
Our framework is able to generate high-resolution textures on multiple data sets. However, there are still several limitations to be considered for future work. 

The largest output that we are able to generate on our testing machine is about 1.6 Gigapixels. Synthesizing and viewing larger images would require the implementation of additional memory management procedures. 

\begin{figure}[t]
	\centering
	\includegraphics[width=\linewidth]{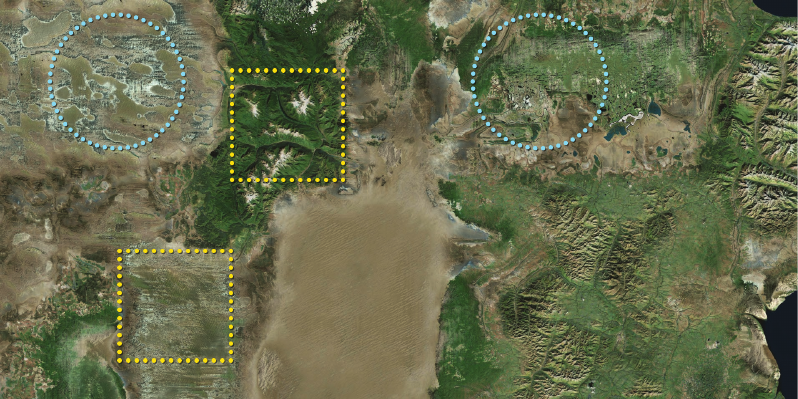}
	\caption{\bd{Artifacts.} When choosing randomized or mismatching neighboring regions, as in this randomized tiling, our textures can exhibit a visible grid-like structure (yellow squares). This effect is more prominent for larger latent tile sizes and can be controlled by choosing appropriate values for the merge level, latent tile size, and MRF.
	Occasionally, the quality of our generated tiles is showing artifacts or unnatural patterns (blue circles). This effect is especially noticeable if multiple defective tiles are selected nearby.}
	\label{fig:artifacts}
\end{figure}

Not all latent tiles generate good looking results, see Fig.~\ref{fig:artifacts} for a failure example where our result contains visual artifacts. Such failures might be attributed to training, sampling of the data set, or inherent limitations of the chosen GAN implementation. Since our framework is fairly modular, we believe that we can integrate new GANs easily.

In addition, the blending of two tiles can lead to unpredictable results. For example, two forest tiles next to each other can generate a visible boundary of different color in the transition region that was not visible in either tile when generated by itself. Fig.~\ref{fig:merge_failure} depicts this failure case, where a merging of apparently visually similar tiles (such as the depicted mixture of arctic ice and continental ice) generates undesirable transition artifacts when attempting to merge them.
Our framework tries to minimize these unpredictable results at the refinement step, but it is not able to eliminate them completely.
Furthermore, such refinement may lead to decreased diversity due to increased repeated similar tiles if the MRF optimization is run till convergence~\cite{Kaspar2015SSTO}. To help avoid outputs with repeated visually similar latent tiles we stop the optimization early. When matching features during the optimization, visual artifacts occurring by directly selecting the highest ranked feature could be reduced by incorporating implicit diversity in the MRF as a regularization loss~\cite{Wang2018Image}. Alternative strategies to the MRF optimization considering latent tile usage \cite{Jamriska2015LazyFluids} or incorporating diversity encouraging feature distance metrics \cite{Mechrez2018TheCL} are worth exploring in future work.

Our results can occasionally exhibit grid-like artifacts, as illustrated in Figure~\ref{fig:artifacts}. This undesirable effect is especially noticeable when initializing a grid tiling completely randomly or when tiling challenging regions where the MRF doesn't manage to select good fitting neighbors and the edge transitions become overly visible. The tiling may also exhibit some repetitiveness when the guidance map contains large regions of uniform color, which are tiled by very similar latent tiles, as visible in the background of Figure~\ref{fig:ui}.

\section{Conclusions and Future Work}
\label{sec:conclusion-future-work}

In this work, we have tackled the problem of texture synthesis in the setting where many input images are given and a large-scale output is required. We have built on recent advances in high-quality generative adversarial networks and proposed a fast algorithm to tile outputs of GANs to produce large plausible texture maps with virtually no boundary artifacts. We have also proposed an interface that enables local and global artistic control on the output image. Our early quantitative and qualitative results demonstrate the fast generation of high-quality textures consisting of hundreds of megapixels. As far as we know, our work is the first to attempt to seamlessly combine intermediate latent tiles at different levels of a GAN to interactively generate such large texture synthesis results. 
 
One interesting venue for future work is to experiment on data\-sets from other celestial bodies (e.g., Mars, Pluto, Sun) and close-ups of everyday objects captured at Gigapixel levels. We are also interested in applying our technique to data with depth information or multiple channels. Another venue for future work is adopting a stacking of multi-layer GANs in order to generate more realistic guide maps that can also be possibly created from an upper layer GAN. Furthermore, the idea of simply manipulating a latent field to produce large textures can be exploited to quickly and consistently modify global appearance including applying color transformation or global patterns. We believe that the idea of latent vector manipulation can lead to many innovations in the future of texture synthesis.

\begin{acks}
We would like to thank Tero Karras and his collaborators~\shortcite{Karras2017ProGAN} for making their source code available. \\
This work was supported by the KAUST Office of Sponsored Research (OSR) under Award No. URF/1/3730-01-01 and URF/1/3426-01-01.
\end{acks}
\bibliographystyle{ACM-Reference-Format}
\bibliography{bibliography}
\end{document}